%% file: twochcircle.tex
\documentclass[twoside]{article}
\newcommand{\mytitle}{A Solvable Two-Charge Ensemble on the Circle} 
\newcommand{\keywords}{circular unitary ensemble, circular orthogonal
  ensemble, random matrix, Pfaffian point process, kernel asymptotics
} 
\newcommand{\msc}{
15B52, %random matrices (algebraic aspects)
60B20, %random matrices (probabilistic aspects)
60G55,  %Point processes
82B23, %statistical mechanics: Exactly solvable models; Bethe ansatz
15A15 %Determinants, permanents, other special matrix functions
}
\input preamble.tex

\newcommand\abs[1]{\left|#1\right|}

\numberwithin{equation}{section} 

\numberwithin{equation}{section}

\usepackage[times]{mathdesign}

\begin{document}
\title{\bfseries\sffamily \mytitle}  
\author{{\sc Christopher Shum} and {\sc Christopher D.~Sinclair}}
\maketitle

\begin{abstract}
  We introduce an ensemble consisting of logarithmically repelling
  charge one and charge two particles on the unit circle
  constrained so that the total charge of all particles equals $N$,
  but the proportion of each species of particle is allowed to vary
  according to a fugacity parameter.   We identify the proper scaling of
  the fugacity with $N$ so that the proportion of each particle stays
  positive in the $N \rightarrow \infty$ limit.  This ensemble forms a
  Pfaffian point process on the unit circle, and we derive the scaling
  limits of the matrix kernel(s) as a function of the interpolating
  parameter. This provides a solvable interpolation between the
  circular unitary and symplectic ensembles.
\end{abstract}

{\bf MSC2010:} \msc

{\bf Keywords:} \keywords
\vspace{1cm}

At their core, classical random matrix ensembles consist of identical
jointly distributed random variables which demonstrate repulsion.
These random variables are often identified with eigenvalues of
matrices chosen with respect to some probability measure (for instance
on the entries of the matrices) or with interacting charged particles
in the complex plane in the presence of a potential, which constrains
the particles to a bounded region.  Three of the simplest examples of
such ensembles are Dyson's circular orthogonal/unitary/symplectic
ensembles C(O/U/S)E, which give rise to random variables with joint
density defined on the $N$-fold copy of the unit circle, $\T^N$, by
\[
\Omega_{N, \beta}(\bs \upzeta) = \frac{1}{Z_{N, \beta}} \prod_{m <
    n}^{N} | \zeta_n - \zeta_m |^{\beta},
\]
where $\beta = 1$ for COE, $\beta = 2$ for CUE and $\beta = 4$ for
CSE, and $Z_{N, \beta}$ is a normalizing constant, the {\em partition
  function} responsible for ensuring that $\Omega_N$ defines a
probability density.  

These specific values of the parameter $\beta$ lead from
well-understood algebraic identities to the expression of the
correlation functions (or joint intensities, defined below) in terms
of determinants (when $\beta = 2$) or Pfaffians (when $\beta = 1, 4$)
of matrices whose entries are determined by a kernel (of the
reproducing sort) dependent on $\beta$ and $N$.  Scaling limits of
these kernels as $N \rightarrow \infty$ then tell us about local
interactions between the random variables as their numbers increase to
infinity.

The special determinantal/Pfaffian structure does not seem to exist
for other values of $\beta$ (see for instance, however
\cite{springerlink:10.1007/s00605-011-0371-8}), so that while it is
perfectly reasonable to generalize the joint density $\Omega_{N,
  \beta}$ to other values of $\beta$, and hence to interpolate between
the classical ensembles, the determinantal/Pfaffian expression for the
correlation functions does not persist along the entirety of this
interpolation.  The goal here is to demonstrate a different
interpolation between COE and CSE, which has Pfaffian intensities for
all values of the interpolating parameter and for which we can
explicitly compute the kernel (as a function of $N$ and the parameter)
and its scaling limits as $N \rightarrow \infty$.

\section{The Two-Charge Model}

The model here is the circular version of the two-charge model
introduced in \cite{rsx}.  We will cover the basics here, but refer
the reader to that article for a more in-depth discussion.  The reader
new to the connection between random matrix theory and two-dimensional
electrostatics should also consult Forrester's book
\cite{forrester-book}.  

We suppose we have $L$ charge one particles
and $M$ charge two particles constrained to the unit circle $\T$ and
interacting logarithmically, so that the interaction between a
particle of charge $q_1$ located at $\zeta_1$ and a particle of charge
$q_2$ located at $\zeta_2$ is given by $-q_1 q_2 \log|z_1 - z_2|$.
Thus, if the $L$ charge one particles are located at $\xi_1, \xi_2,
\ldots, \xi_L$ and the $M$ charge two particles at $\zeta_1, \zeta_2,
\ldots, \zeta_M$, the total interaction energy of the system is given
by
\[
E_{L, M}(\bs \upxi, \bs \upzeta) := -\sum_{\ell < k} \log| \xi_k - \xi_{\ell} |
- 4 \sum_{m < n} \log| \zeta_n - \zeta_m | - 2 \sum_{m=1}^M
\sum_{\ell=1}^L \log|\zeta_m - \xi_{\ell}|.
\]
When the system is at fixed temperature $T$, the probability (density)
of finding the system in the state determined by $(\bs \upxi, \bs
\upzeta)$ is given by
\[
\Omega_{L,M}(\bs \upxi, \bs \upzeta) = \frac{1}{L! M! Z_{L,M}}
e^{-\frac{1}{k T} E_{L,M}(\bs \upxi, \bs \upzeta)},
\]
where $Z_{L,M}$ is a normalizing constant and $k$ is a constant with
units so that $b := (k T)^{-1}$ is a unit-less temperature
parameter\footnote{This parameter is usually denoted $\beta$, but that
  symbol is already taken by the parameter interpolating
  between COE/CUE/CSE.}.  We will assume throughout that $b = 1$. 

We now suppose the number of each species are random variables
constrained so that the sum of all charges is $N$, and the probability of
the system having $L$ charge one and $M$ charge two particles is given by
\[
X^L \frac{Z_{L,M}}{Z_N(X)}
\]
for positive {\em fugacity} parameter $X$, where
\[
Z_N(X) = \sum_{L + 2M = N} \frac{X^L}{L! M!} \int_{\T^L} \int_{\T^M}
e^{-b E_{L,M}(\bs \upxi, \bs \upzeta)} \, d\mu^L(\bs \upxi) \,
d\mu^M(\bs \upzeta),
\]  
and $\mu$ and $\mu^L$ are Lebesgue (Haar) measure\footnote{The
  normalization of Haar measure is unimportant as long as it is done
  consistently.  Here we will take $\mu(\T) = 2 \pi$.} on $\T$ and $\T^L$.  Under this paradigm, our system
consists of two species of particles with total charge summing to $N$,
and $X$ being a parameter which controls the proportion of each
species of particle.

\begin{thm}
\label{thm:1}
If $N$ is even
\[
Z_N(X) = (2 \pi)^{\lfloor \frac{N+1}2 \rfloor} \prod_{n=1}^{\lfloor \frac{N}2
  \rfloor} \frac{(2X)^2 + (N - 2n + 1)^2}{N - 2n + 1},
\]
and if $N$ is odd,
\[
Z_N(X) = (2 \pi)^{\lfloor \frac{N+1}2 \rfloor} X \prod_{n=1}^{\lfloor \frac{N}2
  \rfloor} \frac{(2X)^2 + (N - 2n + 1)^2}{N - 2n + 1}.
\]
\end{thm}

We will often have occasion to compare various particle statistics of
our system with the two-charge model considered in \cite{rsx}.  In
that model (which we will call the RSX model),
the charged particles are restricted to the line (identified with
$\R$) and in the presence of the harmonic oscillator potential---a
potential which keeps the repelling particles from fleeing to
infinity.  
The authors of that work concentrated on global statistics:
the expected number/proportion of each species of particles and the
spatial density for various scalings of the fugacity in the large $N$
limit.  In our case, the limiting expected number/proportion of
particles is certainly of interest (not the least because it requires
a different scaling of the fugacity in order to ensure a positive
proportion of each species of particle).  On the other hand, the
spatial density in our circular model is trivial, since symmetry
demands each species be uniformly distributed on $\T$,
independent of $N$.

The analogy between the circular two charge model here and the RSX
model extends the analogy between the Gaussian Hermitian ensembles,
G(O/S)E and C(O/S)E.  And, as it was for Dyson, the extreme symmetry
present in the circular two-charge ensemble simplifies the derivation
of certain quantities of interest.  For instance, the symmetry of the
circle will allow us to solve for the fluctuations\footnote{Those
  which follow from the scaling limits for all matrix kernels.} in
spacings between particles; this problem is still unsolved in the RSX
model, though universality, suggests that the fluctuations/local
statistics of particles reported here will be the same as in the RSX
model.  Moreover, when the fugacity suitably scales with $N$ we see an
explicit interpolation of matrix kernels between the limiting bulk
kernels for $\beta = 1$ ({\em i.e.} GOE, COE) and $\beta = 4$ (GSE,
CSE) ensembles.

\section{Global Statistics}

If we denote by $L_N(X)$ the random variable giving the number of
charge 1 particles when the fugacity is $X$ and the total charge of
the system is $N$, then $Z_N(T X)/Z_N(X)$, as a function of $T$ is the
probability generating function for $L_N(X)$.  This allows us to
recover the limiting (large $N$) probability generating function for
the number of charge one particles as a function of $X$.  Since the
number of charge 1 particles has the same parity as $N$, we expect
different answers for $N$ even and odd.  Taking this into account, one
sees that in the limit, the number of charge 1 particles is
essentially Poisson. 
\begin{cor}
\label{cor:1}
The limiting probability generating function for the number of charge
1 particles is, when $N$ is even,
\[
\lim_{N \rightarrow \infty} \frac{Z_N(T X)}{Z_N(X)} = \frac{\cosh(\pi
  X T)}{\cosh(\pi 
  X)},
\]
and when $N$ is odd,
\[
\lim_{N \rightarrow \infty} \frac{Z_N(T X)}{Z_N(X)} = \frac{\sinh(\pi
  X T)}{\sinh(\pi X)}.
\]
\end{cor}

From this limiting probability generating function we see that, in the
limit, the expected number of charge 1 particles is finite, though the
exact expectation is dependent on $X$, and (restricting ourselves to
the even case for the moment) is explicitly given by $\pi X \tanh(\pi
X)$.  Thus, if we are interested in a limiting situation with a
non-trivial proportion of charge 1 particles, we need to scale $X$
with $N$.  This calculation suggests the proper scaling of $X$
necessary to achieve this goal is linear; that is $X = N r$.
\begin{thm}
\label{thm:2}
For $r > 0$, 
\[
\lim_{N \rightarrow \infty} \frac{1}{N} E[ L_N(N r)] = 2 r \arctan\left(\frac{1}{2r}\right),
\]
and 
\[
\lim_{N \rightarrow \infty} \frac{1}{N} \mathrm{var}(L_N(N r)) = 2 r
\arctan\left(\frac{1}{2r}\right) - \frac{4 r^2}{1 + 4 r^2}.
\]
\end{thm}

It is interesting to note that the scaling necessary to simultaneously
achieve a non-trivial proportion of each species is different here than
that for the RSX model.  In the RSX case, the fugacity must scale with
$\sqrt{N}$ to achieve this sort of equilibrium.

Unsurprisingly, we find a Central Limit Theorem for the number of
charge 1 particles.
\begin{thm}
\label{thm:4}
For $r > 0$, set
\[
\mu_N = 2 N r \arctan\left(\frac{1}{2r}\right)
\]
and
\[
\sigma_N^2 = N \left(  2 r
\arctan\left(\frac{1}{2r}\right) - \frac{4 r^2}{1 + 4 r^2} \right),
\]
then 
\[
\frac{L_N(N r) - \mu_N}{\sigma_N}
\]
converges in distribution to a standard normal random variable. 
\end{thm}

\section{Local Statistics}

Given a measurable set $A \subseteq \T$ we define the integer valued
random variables $N^{(1)}_A$ and $N^{(2)}_A$ to be respectively the
number of charge 1 and charge 2 particles in $A$.  Of course
$N_A^{(1)}$ and $N_A^{(2)}$ depend also on $X$, but we leave that
dependence implicit.  If for collections of mutually disjoint sets
$A_1, A_2, \ldots, A_{\ell}$ and $B_1, B_2, \ldots, B_m$ there exists
a function $R_{\ell, m} : \T^{\ell} \times \T^m \rightarrow
[0,\infty)$ such that
\[
E\bigg[ \prod_{j=1}^{\ell} N_{A_j}^{(1)} \prod_{k=1}^m N_{B_k}^{(2)}
\bigg] = \int_{\T^{\ell}} \int_{\T^m} R_{\ell, m}(\mathbf x, \mathbf
z) \, d\mu^{\ell}(\mathbf x) \, d\mu^{m}(\mathbf z),
\]
then we call $R_{\ell, m}$ the $(\ell, m)$-{\em intensity} or {\em
  correlation} function.  $R_{\ell, m}$ is dependent on $N$ and $X$, but
we will leave that dependence implicit. Correlation functions are
important, for, for instance, computing the probability that a given
set contains no particles of a certain species.

Like the RSX model, and by essentially the same proof, the correlation
functions can be expressed as the Pfaffian of a $2(\ell + m)$ square
antisymmetric matrix whose entries are given in terms of $2 \times 2$
matrix kernels which encode information about the interactions between
and amongst the different species of particles.
\begin{thm}
\label{thm:5}
There exist matrix kernels $\mathbf K_N^{1,1}, \mathbf K_N^{2,2},
\mathbf K_N^{1,2}, \mathbf K_N^{2, 1} : \T \times \T \rightarrow \C^{2
  \times 2}$ such that
\[
R_{\ell, m}(\mathbf x, \mathbf z) = \Pf \begin{bmatrix}
\left[ \mathbf K_N^{1,1}(x_i, x_j) \right]_{i,j=1}^{\ell} & \left[
  \mathbf K_N^{1,2}(x_i, z_n) \right]_{i,n=1}^{\ell, m} \\
 \left[
  \mathbf K_N^{2,1}(z_k, x_j) \right]_{k,j=1}^{m, \ell} & 
 \left[
  \mathbf K_N^{2,2}(z_k, z_n) \right]_{k,n=1}^{m}
\end{bmatrix}
\]
\end{thm}

Using notation which has now become standard, each of these matrix
kernels can be written as 
\begin{equation}
\label{eq:1}
\mathbf K_N^{s, t}(e^{i \theta}, e^{i \psi}) = \begin{bmatrix}
DS_N^{s, t}(X; \theta, \psi) & S_N^{s, t}(X; \theta, \psi)  \\
-S_N^{t, s}(X; \psi, \theta) & IS_N^{s, t}(X; \theta, \psi)
\end{bmatrix}; \qquad s,t \in \{1,2\},
\end{equation}
where each of the entries is a function $[-\pi, \pi) \times [-\pi, \pi)
\rightarrow \C$.  Exact formulas for the matrix kernels and their
entries will be reported in a subsequent section
(Theorem~\ref{thm:6}).  To make the $X$ dependence explicit, we will write 
$\mathbf K_N^{s,t}(X; \theta, \psi)$ for the right hand side of
(\ref{eq:1}). 

Since the total charge of the system is $N$, we expect that each arc
of length $2 \pi/ N$ will, on average, carry unit charge.  Thus, in
order to investigate the local behavior between particles we place
ourselves on a scale of length $O(N^{-1})$.  That is, the local
statistics, in a neighborhood of $e^{i \varphi}$ are determined by the
scaled kernels
\[
\mathbf K^{s,t}_N\left(X; \varphi + \frac{2 \pi \theta}{N}, \varphi +
  \frac{2 \pi \psi}{N}\right).
\]
For purely geometric reasons, the local kernels must be independent of
$\varphi$, and hence the local statistics are completely determined by
$\mathbf K_N^{s,t}(X; 2 \pi \theta/N, 2\pi \psi/N)$.  We wish to
investigate the large $N$ limit of these kernels, suitably normalized
so that the limit exists.  In order to see non-trivial interactions
between the two species of particles, we need to scale the fugacity
$X$ so that there are $O(N)$ particles of each species.  By
Theorem~\ref{thm:2}, this goal is met when $X = N r$ for fixed $r$,
and hence we define the scaling limits of our kernels to be 
\[
\mathbf K^{s,t}(r; \theta, \psi) = \lim_{N \rightarrow \infty}
\frac{1}{N} \mathbf K^{s,t}_N\left(N r; \frac{2 \pi
    \theta}{N}, \frac{2 \pi \psi}{N}\right).
\]
We define the entries of the scaled kernels to be $DS^{s,t}(r, \theta,
\psi), S^{s,t}(r, \theta, \psi)$ and $IS^{s,t}(r, \theta, \psi)$,
where for instance
\[
S^{s,t}(r; \theta, \psi) = \lim_{N \rightarrow \infty}
\frac{1}{N} S^{s,t}_N\left(N r; \frac{2 \pi
    \theta}{N}, \frac{2 \pi \psi}{N}\right).
\]
Our main result is the evaluation of $\mathbf K^{s,t}(r, \theta,
\psi)$ and the observation that as $r \rightarrow 0^+$ the resulting
Pfaffian point process collapses to that of the Circular Symplectic
(and hence Gaussian Symplectic) ensemble, while as $r \rightarrow
\infty$ we recover the kernel for the Circular (and Gaussian)
Orthogonal Ensemble.  This provides a solvable interpolation
between these ensembles, and by universality should provide for the
limiting local statistics in the RSX ensemble as well.  

\begin{thm}
\label{thm:3}
The entries of $\mathbf K^{1,1}(r; \theta, \psi)$ are given by
\begin{itemize}
\item ${\displaystyle 
S^{1,1}(r; \theta, \psi) = 4 r^2 \int_0^1 \frac{\cos\left( \pi(\theta
    - \psi) t \right)}{4 r^2 + t^2} \, dt
}$
\item ${\displaystyle 
DS^{1,1}(r; \theta, \psi) = i r^2 \int_0^1 \frac{t \sin\left( \pi(\theta
    - \psi) t \right)}{4 r^2 + t^2} \, dt
}$
\item ${\displaystyle 
IS^{1,1}(r; \theta, \psi) = -16 i r^2 \int_0^1 \frac{\sin\left( \pi(\theta
    - \psi) t \right)}{4 r^2 t + t^3} \, dt + 2 \pi \sgn(\psi - \theta)
}$
\end{itemize}
The entries of $\mathbf K^{2,2}(r; \theta, \psi)$ are given by
\begin{itemize}
\item ${\displaystyle 
S^{2,2}(r; \theta, \psi) = \frac{1}{2} \int_0^1 \frac{t^2 \cos\left( \pi(\theta
    - \psi) t \right)}{4 r^2 + t^2} \, dt 
}$
\item ${\displaystyle DS^{2,2}(r; \theta, \psi) = \frac{1}{r^2} DS^{1,1}(r; \theta,
  \psi) = i \int_0^1 \frac{t \sin\left( \pi(\theta
    - \psi) t \right)}{4 r^2 + t^2} \, dt }$
\item ${\displaystyle 
IS^{2,2}(r; \theta, \psi) = -\frac{i}{4} \int_0^1 \frac{t^3 \sin\left( \pi(\theta
    - \psi) t \right)}{4 r^2 + t^2} \, dt 
}$
\end{itemize}
The entries of $\mathbf K^{1,2}(r; \theta, \psi)$ and $\mathbf
K^{2,1}(r; \theta, \psi) = -\mathbf K^{1,2}(r; \theta,
\psi)^{\transpose}$ are given by
\begin{itemize}
\item ${\displaystyle
S^{1,2}(r; \theta, \psi) = r S^{2,2}(r; \theta, \psi) = \frac{r}{2}
\int_0^1 \frac{t^2 \cos\left( \pi(\theta - \psi) t \right)}{4 r^2 + t^2} \, dt  
}$
\item ${\displaystyle
S^{2,1}(r; \theta, \psi) = \frac{1}{r} S^{1,1}(r; \theta, \psi) = 4 r
\int_0^1 \frac{\cos\left( \pi(\theta - \psi) t \right)}{4 r^2 + t^2} \, dt 
}$
\item ${\displaystyle
DS^{1,2}(r; \theta, \psi) = \frac{1}{r} DS^{1,1}(r; \theta, \psi) = i r^2 \int_0^1 \frac{t \sin\left( \pi(\theta
    - \psi) t \right)}{4 r^2 + t^2} \, dt 
}$
\item ${\displaystyle IS^{1,2}(r; \theta, \psi) = -\frac{2}{r}
    DS^{1,1}(r; \theta, \psi) = - 2i r^2 \int_0^1 \frac{t \sin\left( \pi(\theta
    - \psi) t \right)}{4 r^2 + t^2} \, dt  
}$
\end{itemize}
\end{thm}

\begin{figure}[h]
\centering
\includegraphics[scale=.5]{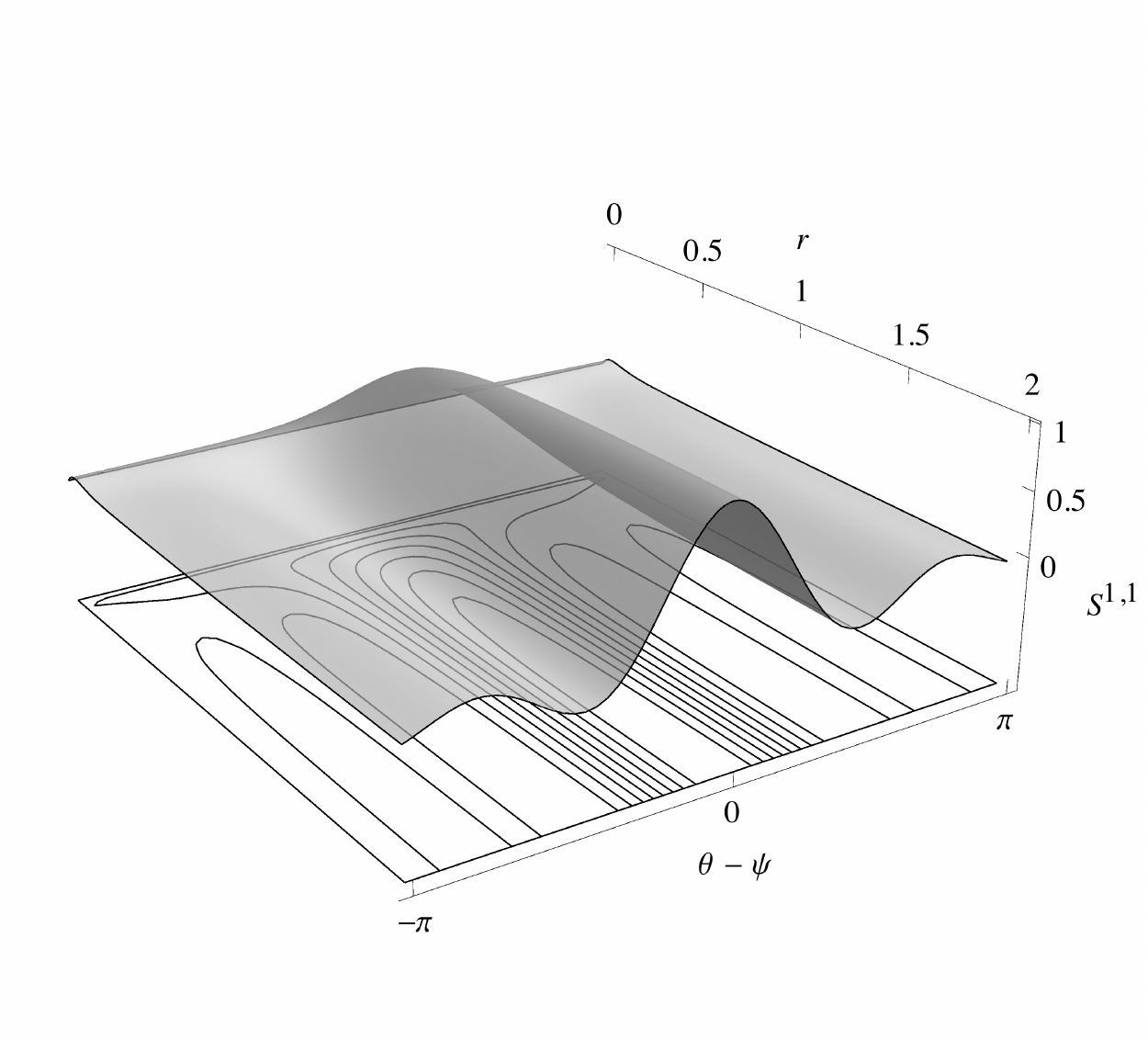}
\includegraphics[scale=.5]{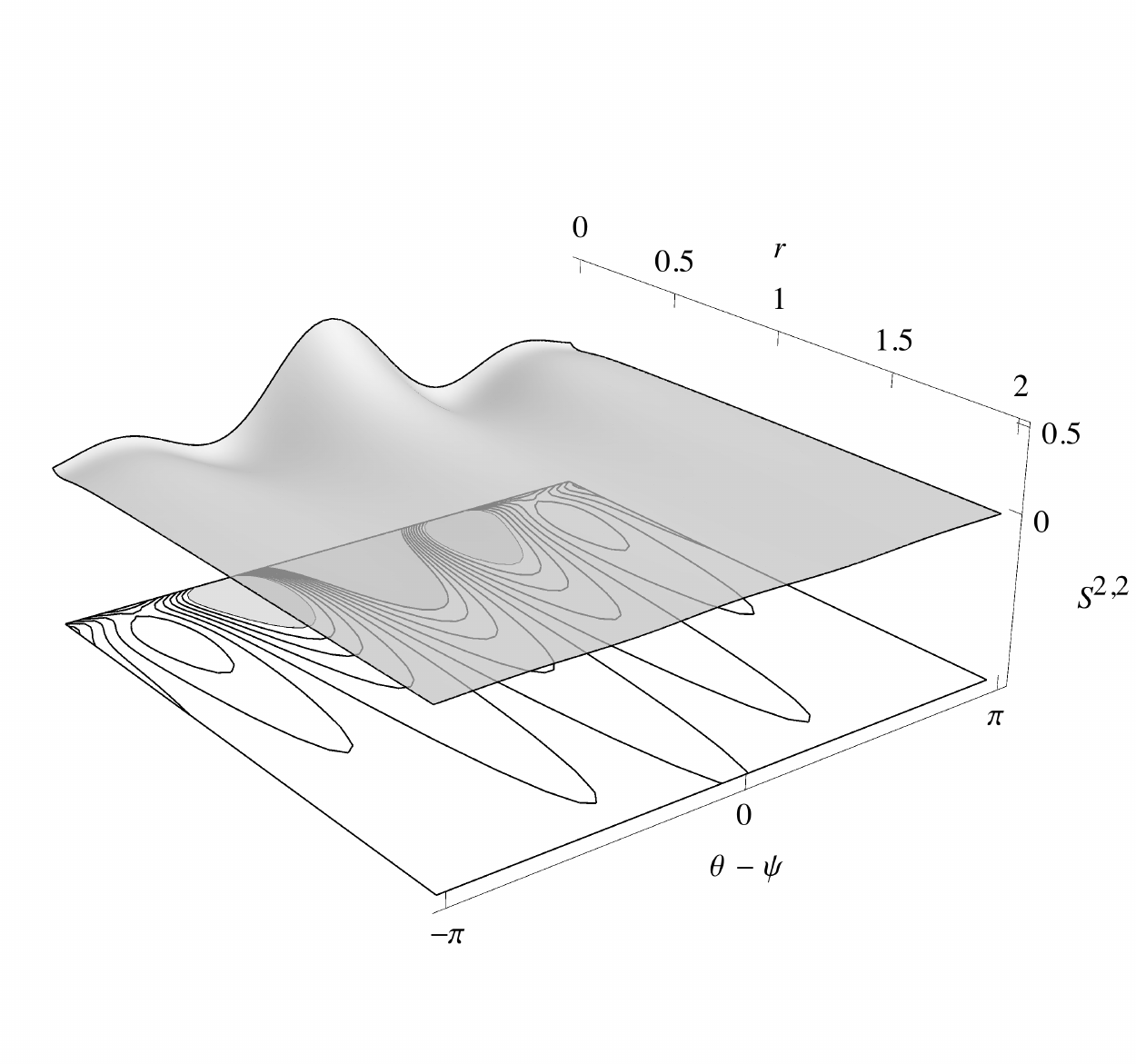}
\begin{caption}{$S^{1,1}$ and $S^{2,2}$ as a function of $r$ and
    $\theta - \psi$.}   
\label{fig:1}
\end{caption}
\end{figure}

As a consistency check, we note that the local spatial density of
charge 1 particles is given by
\[
S^{1,1}(r; \theta, \theta) = 4 r^2 \int_0^1 \frac{1}{4 r^2 + t^2} \,
dt = 2 r \arctan\left( \frac{1}{2r} \right),
\]
which agrees with Theorem~\ref{thm:2}.  Similarly, the local spatial
density of charge 2 particles is given by
\[
S^{2,2}(r; \theta, \theta) = \frac{1}{2} \int_0^1 \frac{t^2}{4 r^2 +
  t^2} \, dt = \frac{1}{2} - r \arctan\left( \frac{1}{2r} \right).
\]
Note that the total local charge density is given by $S^{1,1}(r;
\theta, \theta) + 2 S^{2,2}(r; \theta, \theta) = 1$ as expected. 

The recovery of the kernels for (C/G)OE  and (C/G)SE is the content of
the following corollary.
\begin{cor}
As $r \rightarrow \infty$,
\begin{itemize}
\item${\displaystyle 
S^{1,1}(r; \theta, \psi) \longrightarrow \frac{\sin\left(\pi(\theta -
    \psi) \right)}{\pi(\theta - \psi)} 
}$
\item ${\displaystyle 
DS^{1,1}(r; \theta, \psi) \longrightarrow \frac{i}{4} \left(
  \frac{\sin\left(\pi(\theta - \psi) \right)}{\pi^2(\theta - \psi)^2}
  -   \frac{\cos\left(\pi(\theta - \psi) \right)}{\pi(\theta -
    \psi)}  
\right)
}$
\item ${\displaystyle 
IS^{1,1}(r; \theta, \psi) \longrightarrow -4 i \int_0^1
\frac{\sin\left( \pi(\theta - \psi) t \right)}{t} \, dt + 2 \pi
\sgn(\psi - \theta) 
}$
\end{itemize}
and all entries of the remaining kernels go to 0. 

As $r \rightarrow 0^+$,
\begin{itemize}
\item ${\displaystyle 
S^{2,2}(r; \theta, \psi) \longrightarrow \frac{\sin\left(\pi(\theta -
    \psi) \right)}{2 \pi(\theta - \psi)} 
}$
\item ${\displaystyle 
DS^{2,2}(r; \theta, \psi) \longrightarrow i \int_0^1
\frac{\sin\left( \pi(\theta - \psi) t \right)}{t} \, dt 
}$
\item ${\displaystyle 
IS^{2,2}(r; \theta, \psi) \longrightarrow -\frac{i}{4} \left(
  \frac{\sin\left(\pi(\theta - \psi) \right)}{\pi^2(\theta - \psi)^2}
  -   \frac{\cos\left(\pi(\theta - \psi) \right)}{\pi(\theta -
    \psi)}  
\right) 
}$
\item ${\displaystyle 
IS^{1,1}(r; \theta, \psi) \longrightarrow  2 \pi \sgn(\psi - \theta)  
}$
\end{itemize}
and all other entries of the remaining kernels go to 0. 
\end{cor}

The fact that the $IS^{1,1}$ term `stays up' in the $r \rightarrow
0^+$ limit might seem, on first inspection, suspicious, since in this
situation we are ({\em a posteriori}) tuning the fugacity so that no
charge $1$ particles appear, and thus we expect the $1,1$ kernel to
vanish.  However, the local statistics (in the form of the scaled
correlations)  are determined not by the kernel itself, but the {\em
  Pfaffian} of a matrix formed by the kernel.  The Pfaffian of such a
matrix is unchanged whether or not the limiting $IS^{1,1}$ term stays
up or not (this is a rank-1 perturbation, which the Pfaffian cannot
detect).  In fact, in this limit, any correlation function which has a
component measuring interaction from charge 1 particles will be 0, and
hence the $1,1$ kernel makes no contribution to the limiting
statistics in this instance, independent of the non-zero limit for
$IS^{1,1}$.  This non-zero entry aside, the kernel entries in both
limiting cases can be seen to be essentially equal to the entries of
the kernels for (C/G)OE and (C/G)SE, that is equal up to minor changes
which do not change the Pfaffians appearing in the limiting correlation
functions, appearing in \cite{MR2129906}.  

\section{Acknowledgments}

After posting this manuscript to the arXiv, I (Sinclair) received the
following email from Peter Forrester: 
\begin{quote}
Dear Chris,

I'm sure that back in 2010 you were aware that I introduced the model
of your recent arXiv posting in two papers published in 1984, and that
in my book I extended this work to the computation of the general
correlation functions. Now that your memory has been refreshed, I hope
that you'll appropriately modify your posting.
\end{quote}
We thank Peter for kindly pointing out his previous work in the area
and are happy to acknowledge the existence of \cite{MR765633},
\cite{MR763797} and \cite[\S6.7]{forrester-book}.  It should be
remarked that, while the model considered here is closely related to
Forrester's model, in our model, the number of each type of particle
is a random variable.  It is unsurprising that, in the large $N$ limit
with fugacity tuned so that there is a positive proportion of each
species, the scaled correlation functions match Forrester's in the
limit where $L$ and $M$ (which are non-random in his model) attain
some fixed ratio.  Of course, the introduction of the fugacity
introduces new and interesting questions (some of which we resolve
here) which do not arise in the model with a fixed number of each
species.

Finally, a history of this manuscript is in order.  This is one
chapter of the first author's (Shum's) Ph.D.~thesis
\cite{shum-thesis}, in which several solvable interpolations between
classical random matrix ensembles are studied.  He proposed and
solved, quite independently, the problems in this manuscript after
reading \cite{rsx}, which gives the analogous ensemble on the line,
and left open the determination of the scaling limits of the kernels
in that case.  Of course, universality suggests the scaled (bulk)
kernels in that case should agree with those presented here (and
apparently previously in \cite{forrester-book}) which was the primary
motivation for the present work.

My role, besides giving advice and encouragement, has been to prepare
the manuscript for publication since Chris Shum has since left
academia.   

\section{Proofs}

We will restrict ourselves to the case where $N$ is even.  The odd $N$
cases are more tedious, and we refer to the first author's
Ph.D. thesis for details \cite{shum-thesis} (see also
\cite{sinclair-2007} \cite{sinclair-2008} \cite{mays-forrester}
\cite{borodin-2008} \cite{borodin-2007} where methods of deriving
quantities for odd $N$ from those for even $N$ are discussed for various
ensembles).  Note that, when $N$ is even, so too is $L = N - 2M$.  

\subsection{The Proof of Theorem~\ref{thm:3}}

The proof of this Theorem follows from standard methods in random
matrix theory.  We will give only the basic details and point the
reader to \cite[\S 4.5]{rsx} (see also \cite{sinclair-2007} on which
the ideas in \cite{rsx} rely).

For each non-negative pair of integers $L,M$ with $L + 2M = N$, 
let $\mathbf V_{L,M}(\bs \upxi, \bs \upzeta)$ be the confluent
Vandermonde matrix
\[
\mathbf V_{L,M}(\bs \upxi, \bs \upzeta) = \begin{bmatrix}
1 & \cdots & 1 & 1 & 0 & \cdots  & 1 & 0\\
\xi_1 &  & \xi_L & \zeta_1 & 1 &  & \zeta_M & 1 \\
\xi_1^2 &  \cdots & \xi_L^2 & \zeta_1^2 & 2 \zeta_1 & \cdots
& \zeta_M^2 & 2 \zeta_M \\ 
 \vdots & \ddots & \vdots &  & \vdots & \ddots & \vdots & \\
\xi_1^{n-1} & \cdots & \xi_L^{n-1} & \zeta_1^{n-1} & (n-1)
\zeta_1^{n-2} & \cdots & \zeta_M^{n-1} & (n-1) \zeta_M^{n-2} \\
 \vdots & \ddots & \vdots &  & \vdots & \ddots & \vdots & \\
\xi_1^{N-1} & \cdots & \xi_L^{N-1} & \zeta_1^{N-1} & (N-1)
\zeta_1^{N-2} & \cdots & \zeta_M^{N-1} & (N-1) \zeta_M^{N-2} \\
\end{bmatrix}.
\]
It is well-known, \cite{Meray}, that
\[
\left| \det \mathbf V_{L,M}(\bs
\upxi, \bs \upzeta) \right| = \prod_{k<\ell} | \xi_{\ell} - \xi_k | \prod_{m <
n} | \zeta_n - \zeta_m |^4 \prod_{\ell=1}^L \prod_{m=1}^M |\zeta_m -
\xi_{\ell}|^2 = e^{-E_{L,M}(\bs \upxi, \bs \upzeta)}.
\]
Moreover, for $\zeta, \xi \in \T$, define $\sgn(\zeta - \xi) :=
\sgn(\Arg(\zeta) - \Arg(\xi))$ for the branch of the argument taking
values in $[-\pi, \pi)$.  Then, 
\[
|\zeta - \xi| = -i (\zeta - \xi) \xi^{-1/2} \zeta^{-1/2} \sgn(\zeta - \xi)
\]
and hence,
\[
e^{-E_{L,M}(\bs \upxi, \bs \upzeta)} = \bigg\{ \prod_{\ell=1}^L
e^{\frac{\pi i}{4}} 
\xi_{\ell}^{\frac{1-N}2} \prod_{m=1}^M
\zeta_m^{2 - N} \prod_{k<\ell} \sgn(\xi_{\ell} - \xi_k) \bigg\}
\det \mathbf V_{L,M}(\bs \upxi, \bs \upzeta)
\]
Setting
\[
d\mu_1(\xi) = e^{\frac{\pi i}{4}} \xi^{\frac{-N-1}2} d\mu(\xi)
\qq{and} d\mu_2(\zeta) = 
\zeta^{-N-1} d\mu(\zeta),
\]
we find 
% If we define $\wt{\mathbf V}(\bs \upxi, \bs \upzeta)$ to be the $N
% \times N$ matrix with $n$th row 
% \[
% \begin{bmatrix}
% -i \xi_1^{\frac{2n - N -1}2} & \cdots & -i \xi_L^{\frac{2n - N-1}2}
% & \zeta_1^{\frac{2n - N-1}2} & \frac{2n - N-1}2 \zeta_1^{\frac{2n-N-1}2} & \cdots
% & \zeta_M^{\frac{2n - N-1}2} & \frac{2n - N-1}2 
% \zeta_M^{\frac{2n - N-1}2}  
% \end{bmatrix},
% \]
% then the multilinearity of the determinant implies
% \[
% e^{-E_{L,M}(\bs \upxi, \bs \upzeta)} = \bigg\{ \prod_{k<\ell}
% \sgn(\xi_{\ell} - \xi_k) \bigg\} \det \wt{\mathbf V}_{L,M}(\bs \upxi,
% \bs \upzeta). 
% \]
\[
Z_N(X) = \sum_{(L,M)} \frac{X^L}{L! M!} \int_{\T^L} \int_{\T^M}
\bigg\{ \prod_{k<\ell} \sgn(\xi_{\ell} - \xi_k) \bigg\} \det \mathbf
V_{L,M}(\bs \upxi, \bs \upzeta) d\mu_1^L(\bs \upxi) \, d\mu_2^M(\bs \upzeta).
\]
Here we direct the reader to \cite[\S 4.5]{rsx}, where by replacing
all instances of $\R$ with $\T$, we obtain
\begin{equation}
\label{eq:2}
Z_N(X) = \Pf\left( X^2 \mathbf A_N + \mathbf B_N \right),
\end{equation}
where 
\[
\mathbf A_N = \left[ \int_{\T} \int_{\T} \xi_1^m
\xi_2^n \, \sgn(\xi_2 - \xi_1) \, d\mu_1(\xi_1) \,
d\mu_2(\xi_2) \right]_{m,n=1}^N
\]
and 
\[
\mathbf B_N = \left[ (m - n) \int_{\T} \zeta^m \zeta^n \,
d\mu_2(\zeta) \right].
\]
It remains to evaluate the integrals appearing in $\mathbf A_N$ and
$\mathbf B_N$ and to evaluate the Pfaffian appearing in (\ref{eq:2}).  

Clearly,
\[
 (m - n) \int_{\T} \zeta^{m + n - N - 1} \,
d\mu(\zeta) = \piecewise{
2 \pi (N - 2n + 1) & \mbox{if $n + m = N+1$;} \\
0 & \mbox{otherwise.}
}
\]
An (only slightly) more involved calculation shows
\[
-i \int_{\T} \int_{\T} \xi_1^{\frac{2m-N-1}2}
\xi_2^{\frac{2n-N-1}2} \, \sgn(\xi_2 - \xi_1) \, d\mu(\xi_1) \,
d\mu(\xi_2) =
\piecewise{
{\displaystyle  \frac{8 \pi}{(N - 2n + 1)}} & \mbox{if $n + m = N+1$;} \\ 
0 & \mbox{otherwise.}
}
\]
It follows that $X^2 \mathbf A_N + \mathbf B_N$ is an antisymmetric
matrix with non-zero entries only on the antidiagonal, and hence
\begin{align*}
\Pf(X^2 \mathbf A_N + \mathbf B_N) &= \prod_{n=1}^{N/2} \left(X^2\frac{8
\pi}{N - 2n+1} + 2\pi(N - 2n+1) \right) \\
&= (2\pi)^{N/2}
\prod_{n=1}^{N/2} \frac{(2X)^2 + (2n-1)^2}{2n-1}.
\end{align*}

To prove Corollary~\ref{cor:1}, note that 
\[
\lim_{N\rightarrow\infty} \frac{Z_N(TX)}{Z_N(X)} = \lim_{N \rightarrow 
\infty}  \prod_{n=1}^{N/2} \left(1 + \frac{(2 TX)^2}{(2n-1)^2} \right) 
 \bigg/ \prod_{n=1}^{N/2}  \left(1 + \frac{(2 X)^2}{(2n-1)^2} \right),
\]
this together with the infinite product formula of cosine,
\[
\cos(x) = \prod_{n=1}^{\infty} \left(1 - \frac{4 x^2}{\pi^2(2n-1)^2}
\right),
\]
produces the result. 

\subsection{The Proof of Theorem~\ref{thm:2}}

The characteristic function and cumulant generating functions of
$L_N(X)$ are given by 
\[
\varphi_N(X; t) = \frac{Z_N(X e^{i t})}{Z_N(X)} = \prod_{n=1}^{N/2}
\frac{(2Xe^{i t})^2 + (2n-1)^2}{(2X)^2 + (2n-1)},
\]
and
\begin{align*}
K_N(X; t) &= \log \varphi_N(X; t) \\ 
&= \sum_{n=1}^{N/2} \log \left[ (2 X e^{i t})^2 + (2n-1)^2 \right] -
\sum_{n=1}^{N/2} \log \left[ (2 X)^2 + (2n-1)^2 \right].
\end{align*}
It follows that
\[
E[L_N(X)] = \frac{1}{i} \frac{d}{dt} K_N(X, t) \bigg|_{t=0} =
\sum_{n=1}^{N/2} \frac{8 X^2}{4 X^2 + (2n-1)^2}.
\]
Thus,
\[
\frac{1}{N} E[L_N(N r)] = \frac{2}{N} \sum_{n=1}^{N/2}
\frac{(2r)^2}{(2r)^2 + \left(\frac{2n-1}{N}\right)^2}.
\]
Note that the latter sum is a Riemann sum, and hence 
\[
\lim_{N \rightarrow \infty} \frac{1}{N} E[L_N(N r)] = \int_0^1
\frac{(2r)^2}{(2r)^2 + t^2} \, dt = 2 r \arctan\left(\frac1{2r}\right).
\]

Similarly, 
\[
\mathrm{var}(L_N(X)) = - \frac{d^2}{d t^2} K_N(X, t) \bigg|_{t=0} =
\sum_{n=1}^{N/2} \left( \frac{4X (2n-1)}{4 X^2 + (2n-1)^2}\right)^2. 
\]
And,
\[
\frac{1}{N} \mathrm{var}(L_N(Nr)) = \frac{4}{N} \sum_{n=1}^{N/2}
\frac{(2 r)^2 \left( \frac{2n-1}N \right)^2}{\left((2r)^2 +
\left(\frac{2n-1}N \right)^2 \right)^2}
\]
which again is a Riemann sum, and hence
\[
\lim_{N \rightarrow \infty} \frac{1}{N} \mathrm{var}(L_N(Nr)) = 2
\int_0^1 \frac{(2 r t)^2}{((2r)^2 + t^2)^2} \, dt = 2 r \arctan\left(
\frac1{2r} \right) - \frac{4 r^2}{1 + 4 r^2}.
\]

\subsection{The Proof of Theorem~\ref{thm:4}}

We take the logarithm of the characteristic function

\[
E \left[ \exp \left(it \frac{L_N(Nr) - \mu_N}{\sigma_N} \right)
\right], 
\]
but first we must define a branch of the logarithm.  Because
$\mu_N/\sigma_N = O(N^{-\frac{1}{2}})$, for fixed $t \in \R$,
there exists $N$ large enough such that $-\pi < t
\mu_N/\sigma_N < \pi,$ and so take the branch to be the
negative real line.  Then for this $N$, 
\begin{align}
& \log \left[ E \left[ \exp \left(it \frac{L_N(Nr) - \mu_N}{\sigma_N}
\right) \right] \right] \nonumber \\ 
& \hspace{3cm} = \log \left[ E \left[ \exp \left(it
\frac{L_N(Nr)}{\sigma_N} \right) 
\right] \right] + \log \left[E \left[\exp \left(-it
\frac{\mu_N}{\sigma_N} \right) \right] \right] \nonumber \\ 
& \hspace{3cm}= \log \left[ \prod_{n=1}^\frac{N}{2} \frac{ \left( 2Nr \exp \left(
\frac{it}{\sigma_N} \right) \right)^2 + (2n-1)^2}{(2Nr)^2 + (2n -
1)^2} \right] - it \frac{\mu_N}{\sigma_N} \nonumber \\ 
& \hspace{3cm} = \sum_{n=1}^\frac{N}{2} \log \left[ \frac{ \left( 2Nr \exp \left(
\frac{it}{\sigma_N} \right) \right)^2 + (2n-1)^2}{(2Nr)^2 + (2n -
1)^2} \right] - it \frac{\mu_N}{\sigma_N} \nonumber \\ 
\label{char function}
& \hspace{3cm} = \sum_{n=1}^\frac{N}{2} \log \left[ \frac{ \exp \left(
\frac{2it}{\sigma_N} \right) + \left( \frac{2n-1}{2Nr} \right)^2}{1 +
\left( \frac{2n-1}{2Nr} \right)^2} \right] - it
\frac{\mu_N}{\sigma_N}. 
\end{align}

We claim that the above expression (\ref{char function}) may be replaced by

\begin{equation}
\label{est char function}
\sum_{n = 1}^\frac{N}{2} \left[ \left( \frac{ \frac{2it}{\sigma_N} +
\frac{1}{2!} \left( \frac{2it}{\sigma_N} \right)^2}{1 + \left(
\frac{2n - 1}{2Nr} \right)^2} \right) - \frac{1}{2}\left( \frac{
\frac{2it}{\sigma_N} + \frac{1}{2!} \left( \frac{2it}{\sigma_N}
\right)^2}{1 + \left( \frac{2n - 1}{2Nr} \right)^2} \right)^2 \right]
- it \frac{\mu_N}{\sigma_N}, 
\end{equation}

i.e.\ (\ref{char function}) and (\ref{est char function}) converge to
the same limit as $N \to \infty$.  To see this, first take the Taylor
expansion in (\ref{char function}) for the logarithm, which is valid
since for large enough $N$, the value inside the logarithm is less
than $2$.  So (\ref{char function}) is equal to 
\begin{align}
&\sum_{n = 1}^\frac{N}{2} \sum_{k = 1}^\infty \frac{(-1)^{k + 1}}{k}
\left( \frac{ \exp \left( \frac{2it}{\sigma_N} \right) + \left(
\frac{2n-1}{2Nr} \right)^2}{1 + \left( \frac{2n-1}{2Nr} \right)^2} - 1
\right)^k - it \frac{\mu_N}{\sigma_N} \nonumber \\ \quad 
\label{new char function} 
& \hspace{3cm} = \sum_{n = 1}^\frac{N}{2} \sum_{k = 1}^\infty \frac{(-1)^{k + 1}}{k}
\left( \frac{ \exp \left( \frac{2it}{\sigma_N} \right) - 1}{1 + \left(
\frac{2n-1}{2Nr} \right)^2} \right)^k - it \frac{\mu_N}{\sigma_N}. 
\end{align}

So the absolute value of the difference between (\ref{new char function}) and (\ref{est char function}) is

\begin{align}
&\left\vert \sum_{n = 1}^\frac{N}{2} \left[ \left( \frac{
\frac{2it}{\sigma_N} + \frac{1}{2!} \left( \frac{2it}{\sigma_N}
\right)^2}{1 + \left( \frac{2n - 1}{2Nr} \right)^2} \right) - \left(
\frac{ \exp \left( \frac{2it}{\sigma_N} \right) - 1}{1 + \left(
\frac{2n - 1}{2Nr} \right)^2} \right) \right] \right. \nonumber \\
\quad 
& \hspace{3cm} - \sum_{n = 1}^\frac{N}{2} \left[ \frac{1}{2} \left( \frac{
\frac{2it}{\sigma_N} + \frac{1}{2!} \left( \frac{2it}{\sigma_N}
\right)^2}{1 + \left( \frac{2n - 1}{2Nr} \right)^2} \right)^2 -
\frac{1}{2} \left( \frac{ \exp \left( \frac{2it}{\sigma_N} \right) -
1}{1 + \left( \frac{2n - 1}{2Nr} \right)^2} \right)^2 \right]
\nonumber \\ 
\label{difference}
& \hspace{6cm} + \left. \sum_{n = 1}^\frac{N}{2} \sum_{k = 3}^\infty \frac{(-1)^{k +
1}}{k} \left( \frac{ \exp \left( \frac{2it}{\sigma_N} \right) - 1}{1 +
\left( \frac{2n - 1}{2Nr} \right)^2} \right)^k \right\vert. 
\end{align}
We will show that the absolute values of the three terms converge to $0$
as $N \to \infty$.  

The third term of (\ref{difference}) is 
\begin{align*}
& \abs{\sum_{n = 1}^\frac{N}{2} \sum_{k = 3}^\infty \frac{(-1)^{k +
1}}{k} \left( \frac{ \exp \left( \frac{2it}{\sigma_N} \right) - 1}{1 +
\left( \frac{2n - 1}{2Nr} \right)^2} \right)^k} \\
&\hspace{3cm} \leq \sum_{n = 1}^\frac{N}{2} \left( \abs{\frac{ \exp
\left( \frac{2it}{\sigma_N} \right) - 1}{1 + \left( \frac{2n - 1}{2Nr}
\right)^2}}^3 \abs{\sum_{k = 1}^\infty \frac{(-1)^{k + 4}}{k + 3}
\left( \frac{ \exp \left( \frac{2it}{\sigma_N} \right) - 1}{1 + \left(
\frac{2n - 1}{2Nr} \right)^2} \right)^k} \right). 
\end{align*}
Using the Taylor expansion for the exponential function, there exists
$A > 0$ such that 
\begin{equation}
\label{a1}
\abs{\frac{ \exp \left( \frac{2it}{\sigma_N} \right) - 1}{1 + \left(
\frac{2n - 1}{2Nr} \right)^2}}^3 \leq \frac{A \abs{t}^3}{\sigma_N^3
\left( 1 + \left( \frac{2n - 1}{2Nr} \right)^2 \right)^3} \leq \frac{A
\abs{t}^3}{\sigma_N^3} = \frac{A \abs{t}^3}{\sigma^3 N^3}. 
\end{equation}
Next,
\begin{align}
\abs{\sum_{k = 1}^\infty \frac{(-1)^{k + 4}}{k + 3} \left( \frac{ \exp
\left( \frac{2it}{\sigma_N} \right) - 1}{1 + \left( \frac{2n - 1}{2Nr}
\right)^2} \right)^k} &\leq \abs{ \log \left[ \frac{ \exp \left(
\frac{2it}{\sigma_N} \right) + \left( \frac{2n - 1}{2Nr} \right)^2}{1
+ \left( \frac{2n - 1}{2Nr} \right)^2} \right] } \nonumber \\
&\leq \abs{ \log \left[ \exp \left( \frac{2it}{\sigma_N}
\right) + \left( \frac{2N - 1}{2Nr} \right)^2 \right] }. 
\label{b1}
\end{align}

Thus (\ref{a1}) and (\ref{b1}) show that

\begin{align*}
 & \sum_{n = 1}^\frac{N}{2} \left( \abs{\frac{ \exp \left(
\frac{2it}{\sigma_N} \right) - 1}{1 + \left( \frac{2n - 1}{2Nr}
\right)^2}}^3 \abs{\sum_{k = 1}^\infty \frac{(-1)^{k + 4}}{k + 3}
\left( \frac{ \exp \left( \frac{2it}{\sigma_N} \right) - 1}{1 + \left(
\frac{2n - 1}{2Nr} \right)^2} \right)^k} \right) \nonumber \\
& \hspace{3cm} \leq \sum_{n = 1}^\frac{N}{2} \frac{A \abs{t}^3}{\sigma^3 N^3} \cdot
\abs{ \log \left[ \exp \left( \frac{2it}{\sigma_N} \right) + \left(
\frac{2N - 1}{2Nr} \right)^2 \right] } \\ 
& \hspace{3cm} = \frac{A \abs{t}^3}{2 \sigma^3 N^2} \cdot \abs{ \log
\left[ \exp \left( \frac{2it}{\sigma_N} \right) + \left( \frac{2N -
1}{2Nr} \right)^2 \right] },
\end{align*}
which converges to 0 as $N \rightarrow \infty$.  

Using the same argument from (\ref{a1}), we have the estimate on the
first term of (\ref{difference}) 
\[
\left\vert \sum_{n = 1}^\frac{N}{2} \left[ \left( \frac{
\frac{2it}{\sigma_N} + \frac{1}{2!} \left( \frac{2it}{\sigma_N}
\right)^2}{1 + \left( \frac{2n - 1}{2Nr} \right)^2} \right) - \left(
\frac{ \exp \left( \frac{2it}{\sigma_N} \right) - 1}{1 + \left(
\frac{2n - 1}{2Nr} \right)^2} \right) \right] \right\vert \leq
\sum_{n = 1}^\frac{N}{2} \frac{A \abs{t}^3}{\sigma^3 N^3} = \frac{A
\abs{t}^3}{2 \sigma^3 N^2},
\]
which also goes to 0 as $N \rightarrow \infty$.  

Finally, the summands of the second term in (\ref{difference}) can be
combined as  
\[
\frac{1}{2} \left[ \frac{ \left( \frac{2it}{\sigma_N} + \frac{1}{2!}
\left( \frac{2it}{\sigma_N} \right)^2 - \exp \left(
\frac{2it}{\sigma_N} \right) + 1 \right)^2 + 2 \left(
\frac{2it}{\sigma_N} + \frac{1}{2!} \left( \frac{2it}{\sigma_N}
\right)^2 \right) \left( \exp \left( \frac{2it}{\sigma_N} \right) - 1
\right)}{\left( 1 + \left( \frac{2n - 1}{2Nr} \right)^2 \right)^2}
\right].  
\]
Again, using the same argument as in (\ref{a1}), there exists $A > 0$
such that the absolute value of the first part of the sum is 
\begin{equation}
\label{a2}
\abs{\frac{ \left( \frac{2it}{\sigma_N} + \frac{1}{2!} \left(
\frac{2it}{\sigma_N} \right)^2 - \exp \left( \frac{2it}{\sigma_N}
\right) + 1 \right)^2}{\left( 1 + \left( \frac{2n - 1}{2Nr} \right)^2
\right)^2}} \leq \frac{A^2t^6}{\sigma^6 N^6 \left( 1 + \left(
\frac{2n - 1}{2Nr} \right)^2 \right)^2} \leq \frac{A^2t^6}{\sigma^6
N^6}, 
\end{equation}
and there exists $B > 0$ such that the absolute value of the second
part of the sum is 
\begin{align}
\abs{\frac{2 \left( \frac{2it}{\sigma_N} + \frac{1}{2!} \left(
\frac{2it}{\sigma_N} \right)^2 \right) \left( \exp \left(
\frac{2it}{\sigma_N} \right) - 1 \right)}{\left( 1 + \left( \frac{2n -
1}{2Nr} \right)^2 \right)^2}} &\leq \abs{\frac{2 \left(
\frac{2it}{\sigma N} + \frac{1}{2!} \left( \frac{2it}{\sigma N}
\right)^2 \right) \left( \frac{2Bit}{\sigma N} \right)}{\left( 1 +
\left( \frac{2n - 1}{2Nr} \right)^2 \right)^2}} \nonumber \\ 
&= \abs{\frac{4Bt^2 \left( 1 + \frac{1}{2!} \left( \frac{2it}{\sigma
N} \right) \right)}{\sigma^2 N^2 \left( 1 + \left( \frac{2n - 1}{2Nr}
\right)^2 \right)^2}} \nonumber \\ 
\label{b2}
&\leq \abs{\frac{4Bt^2 \left( 1 + \frac{it}{\sigma N} \right)}{\sigma^2 N^2}}.
\end{align}
Thus, using (\ref{a2}) and (\ref{b2}) the absolute value of the second summand
\begin{align*} 
&\abs{\sum_{n = 1}^\frac{N}{2} \left[ \frac{1}{2}\left( \frac{
\frac{2it}{\sigma_N} + \frac{1}{2!} \left( \frac{2it}{\sigma_N}
\right)^2}{1 + \left( \frac{2n - 1}{2Nr} \right)^2} \right)^2 -
\frac{1}{2} \left( \frac{ \exp \left( \frac{2it}{\sigma_N} \right) -
1}{1 + \left( \frac{2n - 1}{2Nr} \right)^2} \right)^2 \right]}
\\
&\leq \frac{1}{2} \sum_{n = 1}^\frac{N}{2} \left(
\frac{A^2t^6}{\sigma^6 N^6} + \abs{\frac{4Bt^2 \left( 1 +
\frac{it}{\sigma N} \right)}{\sigma^2 N^2}} \right) \\
&= \frac{1}{2 \sigma^2 N^2} \sum_{n = 1}^\frac{N}{2} \left(
\frac{A^2t^6}{\sigma^4 N^4} + \abs{4Bt^2 \left( 1 + \frac{it}{\sigma
N} \right)} \right) \\
&= \frac{1}{4 \sigma^2 N} \left( \frac{A^2t^6}{\sigma^4 N^4} +
\abs{4Bt^2 \left( 1 + \frac{it}{\sigma N} \right)} \right) \\
\end{align*}
which converges to 0 as $N \rightarrow \infty$. 
Thus, (\ref{difference}) converges to $0$, so (\ref{char function})
and (\ref{est char function}) converge to the same limit. 

We rewrite (\ref{est char function}) as 

\begin{align*}
& \sum_{n = 1}^\frac{N}{2} \left[ \left( \frac{ \frac{2it}{\sigma_N} +
\frac{1}{2!} \left( \frac{2it}{\sigma_N} \right)^2}{1 + \left(
\frac{2n - 1}{2Nr} \right)^2} \right) - \frac{1}{2}\left( \frac{
\frac{2it}{\sigma_N} + \frac{1}{2!} \left( \frac{2it}{\sigma_N}
\right)^2}{1 + \left( \frac{2n - 1}{2Nr} \right)^2} \right)^2 \right]
- it \frac{\mu_N}{\sigma_N} \\
& \hspace{1cm} = \underbrace{\sum_{n = 1}^\frac{N}{2} \left[
    -\frac{2t^2}{\sigma_N^2 \left(1 + \left( \frac{2n - 1}{2Nr}
        \right)^2 \right)} + \frac{2t^2}{\sigma_N^2 \left( 1 + \left(
          \frac{2n - 1}{2Nr} \right)^2
      \right)^2} \right]}_{(*)} \\
& \hspace{1cm} + \underbrace{\sum_{n = 1}^\frac{N}{2} \left[
    \frac{2it}{\sigma_N \left(1 + \left( \frac{2n - 1}{2Nr} \right)^2
      \right)} + \frac{4it^3}{\sigma_N^3 \left(1 + \left( \frac{2n -
            1}{2Nr} \right)^2 \right)^2} - \frac{2t^4}{\sigma_N^4
      \left(1 + \left( \frac{2n - 1}{2Nr} \right)^2 \right)^2} \right]
  - it \frac{\mu_N}{\sigma_N}}_{(**)}.
\end{align*}

We will show that the expressions $(*)$ and $(**)$ converge to
$-\frac{t^2}{2}$ and $0$, respectively.  First we define 
\[
\mu = \frac{\mu_N}{N} = 2r \arctan\left( \frac{1}{2r} \right),
\]
and let $\sigma$ be the positive number such that
\[
\sigma^2 = \frac{\sigma^2_N}{N} = \left( 2r\arctan\left( \frac{1}{2r}
\right) - \frac{4r^2}{1 + 4r^2} \right). 
\]
Then,
\begin{align*}
-\frac{2t^2}{\sigma_N^2} \sum_{n = 1}^\frac{N}{2} \frac{1}{1 + \left(
\frac{2n - 1}{2Nr} \right)^2} &= -\frac{t^2}{\sigma^2} \cdot
\frac{2}{N} \sum_{n = 1}^\frac{N}{2} \frac{1}{1 + \left( \frac{2n -
1}{2Nr} \right)^2} \\
&\longrightarrow -\frac{t^2}{\sigma^2} \int_0^1 \frac{1}{1 + \left( \frac{t}{2r}
\right)^2 } \: dt = -2r \arctan \left( \frac{1}{2r} \right) \cdot
\frac{t^2}{\sigma^2},
\end{align*}
and
\begin{align*}
\frac{2t^2}{\sigma_N^2} \sum_{n = 1}^\frac{N}{2} \frac{1}{ \left( 1 +
\left( \frac{2n - 1}{2Nr} \right)^2 \right)^2} &=
\frac{t^2}{\sigma^2} \cdot \frac{2}{N} \sum_{n = 1}^\frac{N}{2}
\frac{1}{ \left( 1 + \left( \frac{2n - 1}{2Nr} \right)^2 \right)^2}
\\
&\longrightarrow \frac{t^2}{\sigma^2} \int_0^1 \frac{1}{ \left( 1 +
\left( \frac{t}{2r} \right)^2 \right)^2 } \: dt  
\\
&= \left( 2r \arctan \left( \frac{1}{2r} \right) + \frac{4r^2}{1 +
4r^2} \right) \frac{t^2}{2\sigma^2}. 
\end{align*}
Thus, changing $N/2 \mapsto N$, 
\begin{align*}
&\sum_{n = 1}^N \left[ -\frac{2t^2}{\sigma_{2N}^2 \left(1 + \left(
        \frac{2n-1}{4Nr} \right)^2 \right)} + 
  \frac{2t^2}{\sigma_{2N}^2 \left( 1 + \left( \frac{2n - 1}{4Nr}
      \right)^2 \right)^2} \right] \\ 
& \hspace{2cm} \longrightarrow \left( 2r \arctan \left( \frac{1}{2r} \right) +
\frac{4r^2}{1 + 4r^2} \right) \frac{t^2}{2\sigma^2} - 2r \arctan 
\left( \frac{1}{2r} \right) \cdot \frac{t^2}{\sigma^2} \\ 
& \hspace{2cm} = -\frac{t^2}{2}. 
\end{align*} 

Next,
\begin{align*}
\frac{2it}{\sigma_N} \sum_{n = 1}^\frac{N}{2} \frac{1}{1 + \left(
    \frac{2n - 1}{2Nr} \right)^2} - it \frac{\mu_N}{\sigma_N} &=
\frac{\sqrt{N}it}{\sigma} \left( \frac{2}{N} \sum_{n = 1}^\frac{N}{2}
  \frac{1}{1 + \left( \frac{2n - 1}{2Nr} \right)^2} - \mu \right)
\\
&= \frac{\sqrt{N}it}{\sigma} \left( \frac{2}{N} \sum_{n =
    1}^\frac{N}{2} \frac{1}{1 + \left( \frac{2n - 1}{2Nr} \right)^2} -
  \int_0^1 \frac{1}{1 + \left( \frac{t}{2r} \right)^2} \: dt
\right).
\end{align*}
It is a well-known calculus fact that the error term of the Riemann
sum using the midpoint rule is $O(N^{-2})$ if the integrand is twice
differentiable.  That is, 
\[
\frac{2it}{\sigma_N} \sum_{n = 1}^\frac{N}{2} \left[ \frac{1}{1 + \left( \frac{2n - 1}{2Nr} \right)^2} \right] - it \frac{\mu_N}{\sigma_N} \longrightarrow 0. 
\]
Similarly,
\[ 
-\frac{2t^4}{\sigma_N^4} \sum_{n = 1}^\frac{N}{2} \frac{1}{\left(1 +
    \left( \frac{2n - 1}{2Nr} \right)^2 \right)^2} =
-\frac{t^4}{N\sigma^4} \cdot \frac{2}{N} \sum_{n = 1}^\frac{N}{2}
\frac{1}{\left(1 + \left( \frac{2n - 1}{2Nr} \right)^2 \right)^2}
\longrightarrow 0
\]
and since the sums converge to finite integrals,
\[
\frac{4it^3}{\sigma_N^3} \sum_{n = 1}^\frac{N}{2} \frac{1}{\left(1 +
    \left( \frac{2n - 1}{2Nr} \right)^2 \right)^2} =
\frac{2it^3}{\sqrt{N}\sigma^3} \cdot \frac{2}{N} \sum_{n =
  1}^\frac{N}{2} \frac{1}{\left(1 + \left( \frac{2n - 1}{2Nr}
    \right)^2 \right)^2} \longrightarrow 0
\]

This shows that
\[
\log \left[ \E \left[ \exp \left(it \frac{L_N(Nr) - \mu_N}{\sigma_N}
    \right) \right] \right]  \longrightarrow -\frac{t^2}{2}, 
\]
establishing the Central Limit Theorem for $L_N(Nr)$.  

\subsection{The Proof of Theorem~\ref{thm:5}}
\label{sec:proof-theor-refthm:5}

Theorem~\ref{thm:5} follows {\em mutadis mutandis} from the proof of
Theorem 3.3 in \cite{rsx} (which is based on \cite[\S7]{borodin-2008},
\cite{sinclair-2008} and ultimately \cite{MR1657844}).  In order to
appeal directly to \cite[\S4.6]{rsx}, we define for indeterminants
$a_1, \ldots, a_N, b_1, \ldots, b_N$ and $x_1, \ldots, x_N, z_1,
\ldots, z_N \in \T$ the measures 
\[
\eta_1(\xi) = \sum_{n=1}^N a_n \delta(\xi - x_n) \qq{and}
\eta_2(\zeta) = \sum_{n=1}^N b_n \delta(\zeta - z_n),
\]
where $\delta(0)$ is probability measure with unit mass at $0$.
From this we define the measures
\[
d\nu_1(\xi) = e^{\frac{\pi i}4} \xi^{\frac{-N-1}2} ( d\mu(\xi) +
d\eta_1(\xi) ) \qq{and} d\nu_2(\zeta) = \zeta^{-N-1} (d\mu(\zeta) +
d\eta_2(\zeta)).
\]
Then, $R_{\ell, m}(x_1, \ldots, x_{\ell}, z_1, \ldots, z_m)$ is the
coefficient of $a_1 \cdots a_{\ell} b_1 \cdots b_m$ in 
\begin{equation}
\label{eq:3}
Z_N^{\nu_1, \nu_2}(X) := \sum_{L + 2M = N} \frac{1}{L! M!} | \det
\mathbf V_{L,M}(\bs \upxi, \bs \upzeta) | \, d\nu_1^L(\bs \upxi) \,
d\nu_2^M(\bs \upzeta).
\end{equation}
Computing the right-hand side of (\ref{eq:3}) as a Pfaffian and
reading off the desired coefficient allows us to express $R_{\ell,
  m}(\mathbf x, \mathbf z)$ as the Pfaffian of a matrix of the form
\[
\Pf \begin{bmatrix}
\left[ \mathbf K_N^{1,1}(x_i, x_j) \right]_{i,j=1}^{\ell} & \left[
  \mathbf K_N^{1,2}(x_i, z_n) \right]_{i,n=1}^{\ell, m} \\
 \left[
  \mathbf K_N^{2,1}(z_k, x_j) \right]_{k,j=1}^{m, \ell} & 
 \left[
  \mathbf K_N^{2,2}(z_k, z_n) \right]_{k,n=1}^{m}
\end{bmatrix}
\]
In fact, the calculation in \cite{rsx} yields the explicit form of
$\mathbf K_N^{1,1}, \mathbf K_N^{2,2}$ and 
$\mathbf K_N^{1,2}$ and $\mathbf K_N^{2,1} = -(\mathbf
K_N^{1,2})^{\transpose}$.  We report the entries of these matrix
kernels here, but leave the details of the calculation to the reader. 
\begin{thm}
\label{thm:6}
Assuming $N$ is even, and with notation as in (\ref{eq:1}), the
entries of $\mathbf K^{1,1}_N(X; \theta, \psi)$ are given
by 
\begin{itemize}
\item ${\displaystyle 
S_N^{1,1}(X; \theta, \psi) = \frac{4 X^2}{\pi} \sum_{n=1}^{N/2} \frac{
  \cos\left( \left(n - \frac12\right) (\theta - \psi) \right)}{(2X)^2
  + (2n-1)^2} 
}$
\item ${\displaystyle 
DS_N^{1,1}(X; \theta, \psi) = \frac{i X^2}{\pi} \sum_{n=1}^{N/2} \frac{(2n-1)
  \sin\left( \left(n - \frac12\right) (\theta - \psi) \right)}{(2X)^2
  + (2n-1)^2}  
}$
\item ${\displaystyle 
IS_N^{1,1}(X; \theta, \psi) = -\frac{16 i X^2}{\pi} \sum_{n=1}^{N/2}    
\frac{
  \sin\left( \left(n - \frac12\right) (\theta - \psi)
  \right)}{(2n-1)\left((2X)^2 + (2n-1)^2\right)} + \sgn(\psi - \theta)
}$
\end{itemize}
The entries of  $\mathbf K^{2,2}_N(X; \theta, \psi)$ are
given by 
\begin{itemize}
\item ${\displaystyle 
S_N^{2,2}(X; \theta, \psi) =  \frac{1}{2 \pi} \sum_{n=1}^{N/2} \frac{(2n-1)^2
  \cos\left( \left(n - \frac12\right) (\theta - \psi) \right)}{(2X)^2
  + (2n-1)^2} 
}$
\item ${\displaystyle 
DS_N^{2,2}(X; \theta, \psi) =  \frac{i}{\pi} \sum_{n=1}^{N/2} \frac{ (2n-1)
  \sin\left( \left(n - \frac12\right) (\theta - \psi) \right)}{(2X)^2
  + (2n-1)^2} 
}$
\item ${\displaystyle 
IS_N^{2,2}(X; \theta, \psi) =  \frac{-i}{4 \pi} \sum_{n=1}^{N/2}
\frac{ (2n-1)^3
  \sin\left( \left(n - \frac12\right) (\theta - \psi) \right)}{(2X)^2
  + (2n-1)^2} 
}$
\end{itemize}
The entries of  $\mathbf K^{1,2}_N(X; \theta, \psi)$ and
$\mathbf K^{2,1}_N(X; \theta, \psi) = - \mathbf
K^{1,2}_N(X; \psi, \theta)^{\transpose}$ are given by 
\begin{itemize}
\item ${\displaystyle 
S_N^{1,2}(X; \theta, \psi) =  \frac{X}{2 \pi} \sum_{n=1}^{N/2} \frac{
(2n-1)^2  \cos\left( \left(n - \frac12\right) (\theta - \psi)
\right)}{(2X)^2 
  + (2n-1)^2} 
}$
\item ${\displaystyle 
S_N^{2,1}(X; \theta, \psi) =  \frac{4 X}{\pi} \sum_{n=1}^{N/2} \frac{
  \cos\left( (2n-1) \left(n - \frac12\right) (\theta - \psi) \right)}{(2X)^2
  + (2n-1)^2} 
}$
\item ${\displaystyle 
DS_N^{1,2}(X; \theta, \psi) =  \frac{i X}{\pi} \sum_{n=1}^{N/2}
\frac{ (2n-1)
\sin\left( \left(n - \frac12\right) (\theta - \psi) \right)}{(2X)^2
  + (2n-1)^2} 
}$
\item ${\displaystyle 
IS_N^{1,2}(X; \theta, \psi) =  -\frac{2 i X}{\pi} \sum_{n=1}^{N/2}
\frac{ (2n-1)
  \sin\left( \left(n - \frac12\right) (\theta - \psi) \right)}{(2X)^2
  + (2n-1)^2} 
}$
\end{itemize}
\end{thm}

\subsection{The Proof of Theorem~\ref{thm:3}}

Write $\mathbf K_N^{(\ell, m)}(X; \bs \uptheta, \bs \uppsi)$ for the $2
\ell + 2 m$ square matrix 
\[
\mathbf K_N^{(\ell, m)}(X; \bs \uptheta, \bs \uppsi)
\begin{bmatrix}
\left[ \mathbf K_N^{1,1}(\theta_i, \theta_j) \right]_{i,j=1}^{\ell} & \left[
  \mathbf K_N^{1,2}(\theta_i, \psi_n) \right]_{i,n=1}^{\ell, m} \\
 \left[
  \mathbf K_N^{2,1}(\psi_k, \theta_j) \right]_{k,j=1}^{m, \ell} & 
 \left[
  \mathbf K_N^{2,2}(\psi_k, \psi_n) \right]_{k,n=1}^{m}
\end{bmatrix} 
\]
The Pfaffian of this matrix, of course, yields the $\ell, m$
correlation function (in terms of the arguments of the variables).  

In order to calculate the large $N$ limit of the correlations when the
fugacity is tuned so that there are a positive proportion of both
species in the limit, we set $X = N r$ for $r > 0$.  Using the
well-known Pfaffian identity,  
\[
\Pf( \mathbf B \mathbf A \mathbf B^{\transpose} ) = \det \mathbf B \Pf
\mathbf A
\]
we may multiply $\mathbf K_N^{(\ell, m)}(X; \bs \uptheta, \bs \uppsi)$
on the left and right by the diagonal matrix
\[
\mathbf D = \mathrm{diag}\Bigg[ \underbrace{\sqrt{\frac{r}{X}},
    \sqrt{\frac{X}{r}}, \cdots, \sqrt{\frac{r}{X}},
    \sqrt{\frac{X}{r}}}_{2 \ell},  \underbrace{\sqrt{\frac{X}{r}},
    \sqrt{\frac{r}{X}}, \cdots, \sqrt{\frac{X}{r}},
    \sqrt{\frac{r}{X}}}_{2 m} \Bigg]
\]
without changing its Pfaffian.  That is, we may replace the kernel
entries given in Theorem~\ref{thm:6} with the following, without
changing the correlation functions.  

It is clear that this procedure changes the kernels in a manner
independent of $\ell$ and $m$.  The new entries of $\mathbf
K_N^{1,1}(X; \theta, \psi)$ are given by
\begin{itemize}
\item ${\displaystyle 
S_N^{1,1}(X; \theta, \psi) = \frac{4 X^2}{\pi} \sum_{n=1}^{N/2} \frac{
  \cos\left( \left(n - \frac12\right) (\theta - \psi) \right)}{(2X)^2
  + (2n-1)^2} 
}$
\item ${\displaystyle 
DS_N^{1,1}(X; \theta, \psi) = \frac{i X r}{\pi} \sum_{n=1}^{N/2} \frac{(2n-1)
  \sin\left( \left(n - \frac12\right) (\theta - \psi) \right)}{(2X)^2
  + (2n-1)^2}  
}$
\item ${\displaystyle 
IS_N^{1,1}(X; \theta, \psi) = -\frac{16 i X^3}{\pi r} \sum_{n=1}^{N/2}    
\frac{
  \sin\left( \left(n - \frac12\right) (\theta - \psi)
  \right)}{(2n-1)\left((2X)^2 + (2n-1)^2\right)} + \frac{X}r \sgn(\psi - \theta)
}$
\end{itemize}
The new entries of  $\mathbf K^{2,2}_N(X; \theta, \psi)$ are
given by 
\begin{itemize}
\item ${\displaystyle 
S_N^{2,2}(X; \theta, \psi) =  \frac{1}{2 \pi} \sum_{n=1}^{N/2} \frac{(2n-1)^2
  \cos\left( \left(n - \frac12\right) (\theta - \psi) \right)}{(2X)^2
  + (2n-1)^2} 
}$
\item ${\displaystyle 
DS_N^{2,2}(X; \theta, \psi) =  \frac{1}{r^2} DS_N^{1,1}(X; \theta, \psi)
}$
\item ${\displaystyle 
IS_N^{2,2}(X; \theta, \psi) =  \frac{-i r}{4 X \pi} \sum_{n=1}^{N/2}
\frac{ (2n-1)^3
  \sin\left( \left(n - \frac12\right) (\theta - \psi) \right)}{(2X)^2
  + (2n-1)^2} 
}$
\end{itemize}
The new entries of  $\mathbf K^{1,2}_N(X; \theta, \psi)$ and
$\mathbf K^{2,1}_N(X; \theta, \psi) = - \mathbf
K^{1,2}_N(X; \psi, \theta)^{\transpose}$ are given by 
\begin{itemize}
\item ${\displaystyle 
S_N^{1,2}(X; \theta, \psi) = r S_N^{2,2}(X; \theta, \psi)
}$
\item ${\displaystyle 
S_N^{2,1}(X; \theta, \psi) = \frac{1}{r} S_N^{1,1}(X; \theta, \psi)
}$
\item ${\displaystyle 
DS_N^{1,2}(X; \theta, \psi) = \frac{1}{r} DS_N^{1,1}(X; \theta, \psi)
}$
\item ${\displaystyle 
IS_N^{1,2}(X; \theta, \psi) =  -\frac{2}{r} DS_N^{1,1}(X; \theta, \psi)
}$
\end{itemize}

Computing the limit 
\begin{align*}
S^{1,1}(r; \theta, \psi) &= \lim_{N \rightarrow \infty} \frac{2
  \pi}{N} S_N^{1,1}\left(Nr; \frac{2 
    \pi \theta}{N}, \frac{2 \pi \psi}{N} \right) \\
&= \lim_{N \rightarrow \infty} 8 N r^2  \sum_{n=1}^{N/2} \frac{
  \cos\left( \frac{\pi}{N} \left(2 n - 1\right) (\theta - \psi)
  \right)}{(2 N r)^2 + (2n-1)^2} \\
&= 4 r^2 \lim_{N \rightarrow \infty} \frac{2}{N}  \sum_{n=1}^{N/2} \frac{
  \cos\left( \pi \left(\frac{2 n - 1}{N}\right) (\theta - \psi)
  \right)}{(2 r)^2 + \left(\frac{2n-1}{N}\right)^2} \\ 
&= 4 r^2 \int_0^1 \frac{\cos\left( \pi(\theta
    - \psi) t \right)}{4 r^2 + t^2} \, dt.
\end{align*}
The calculation of the analogous limits defining $DS^{1,1}$, $IS^{1,1},
S^{2,2}$ and $IS^{2,2}$ are all similar to that for $S^{1,1}$, 
and left to the reader.

\bibliography{bibliography}

\begin{center}
\noindent\rule{4cm}{.5pt}
\vspace{.25cm}

\noindent {\sc \small Christopher Shum}\\
\noindent {\sc \small Christopher D.~Sinclair}\\
{\small Department of Mathematics, University of Oregon, Eugene OR 97403} \\
corresponding email: {\tt csinclai@uoregon.edu}
\end{center}

\end{document}

%% file: preamble.tex
\usepackage{amsmath,amssymb,
  amsthm,amsfonts,amscd,mathrsfs,pifont,upgreek} 
\usepackage{eucal}  % Euler fonts
\usepackage{verbatim}
\usepackage{ifpdf}
\usepackage[colorlinks=true, citecolor=blue]{hyperref}

\ifpdf
        \usepackage[pdftex]{graphicx}
\else
        \usepackage[dvips]{graphicx}
\fi
\newtheorem{thm}{Theorem}[section]
\newtheorem{cor}[thm]{Corollary}

\newtheorem{thm*}{Theorem}[]
\newtheorem{cor*}[thm*]{Corollary}
\newtheorem{claim*}[thm*]{Claim}
\newtheorem{lemma*}[thm*]{Lemma}
\newtheorem{prop*}[thm*]{Proposition}
\newtheorem{conj*}[thm*]{Conjecture}

\theoremstyle{definition}

\newtheorem{problem*}{Problem}[section]

\newtheorem{question*}{Question}[section]

\newtheorem{defn*}{Definition}

\theoremstyle{remark}

\newcommand{\qq}[1]{\qquad \mbox{#1} \qquad}

\newcommand{\BB}[1]{\ensuremath{\mathbb{#1}}}
\newcommand{\E}{\ensuremath{\BB{E}}}

\newcommand{\R}{\ensuremath{\BB{R}}}

\newcommand{\C}{\ensuremath{\BB{C}}}
\newcommand{\T}{\ensuremath{\BB{T}}}

\newcommand{\bs}{\ensuremath{\boldsymbol}}

\newcommand{\Arg}{\ensuremath{\mathrm{Arg}}}

\newcommand{\transpose}{\ensuremath{\mathsf{T}}}

\newcommand{\piecewise}[1]{
\left\{
\begin{array}{ll}
#1
\end{array}
\right.
}

\DeclareMathOperator{\sgn}{sgn}

\DeclareMathOperator{\Pf}{Pf}